\documentclass[11pt,onecolumn]{article}
\usepackage[T1]{fontenc}
\usepackage[latin9]{inputenc}
\usepackage{amsmath}
\usepackage{graphicx}
\usepackage{float}
\usepackage{authblk}

\usepackage{lmodern}
\usepackage[T1]{fontenc}
\usepackage[latin9]{inputenc}
\usepackage[unicode=true,
 bookmarks=false,
breaklinks=false,pdfborder={0 0 1},backref=false,colorlinks=false]
 {hyperref}
\usepackage[a4paper, total={210mm,297mm},margin=2.0cm]{geometry}
\usepackage{breakurl}
\usepackage[normalem]{ulem}
\usepackage{color}

\usepackage{datetime}
\newdate{date}{03}{10}{2017}

\usepackage[english]{babel}

\providecommand{\tabularnewline}{\\}

\title{Effects of Nanoparticles on the Dynamic Morphology of Electrified Jets}

\author[1]{Marco Lauricella \thanks{Electronic address: \texttt{m.lauricella@iac.cnr.it}; Corresponding author}}
\author[2,3]{Dario Pisignano}
\author[1,4]{Sauro Succi}

\affil[1]{Istituto per le Applicazioni del Calcolo CNR, Via dei Taurini 19, 00185 Rome, Italy}
\affil[2]{Dipartimento di Matematica e Fisica \textquotedbl{}Ennio De Giorgi\textquotedbl{}, University of Salento, via Arnesano, 73100 Lecce, Italy}
\affil[3]{NEST, Istituto Nanoscienze-CNR, Piazza S. Silvestro 12, 56127 Pisa, Italy}
\affil[4]{Institute for Applied Computational Science, Harvard John A. Paulson School of Engineering And Applied Sciences, Cambridge, MA 02138, United States}

\date{\displaydate{date}}

\begin{document}

\maketitle

\begin{abstract}
We investigate the effects of nanoparticles on the onset of varicose
and whipping instabilities in the dynamics of electrified jets. 
In particular, we show that the non-linear interplay between the mass
of the nanoparticles and electrostatic instabilities, gives rise to qualitative
changes of the dynamic morphology of the jet, which in turn, drastically
affect the final deposition pattern in electrospinning experiments. 
It is also shown that even a tiny amount of excess mass, of the order
of a few percent, may more than double the radius of the 
electrospun fiber, with substantial implications for the design of
experiments involving electrified jets as well as spun organic fibers.
\end{abstract}

Dynamic instabilities of charged polymeric liquid jets play a crucial role
in many natural phenomena and industrial processes, such as electrospinning, 
ink-jet printing, electrospray, and many others 
\cite{andradyscitechnol2008, yarin2014fundamentals, pisignano2013polymer,hohman2001electrospinning,eggers2008physics}.

\begin{figure}[!ht]
\begin{centering}
\includegraphics[width=0.46\textwidth]{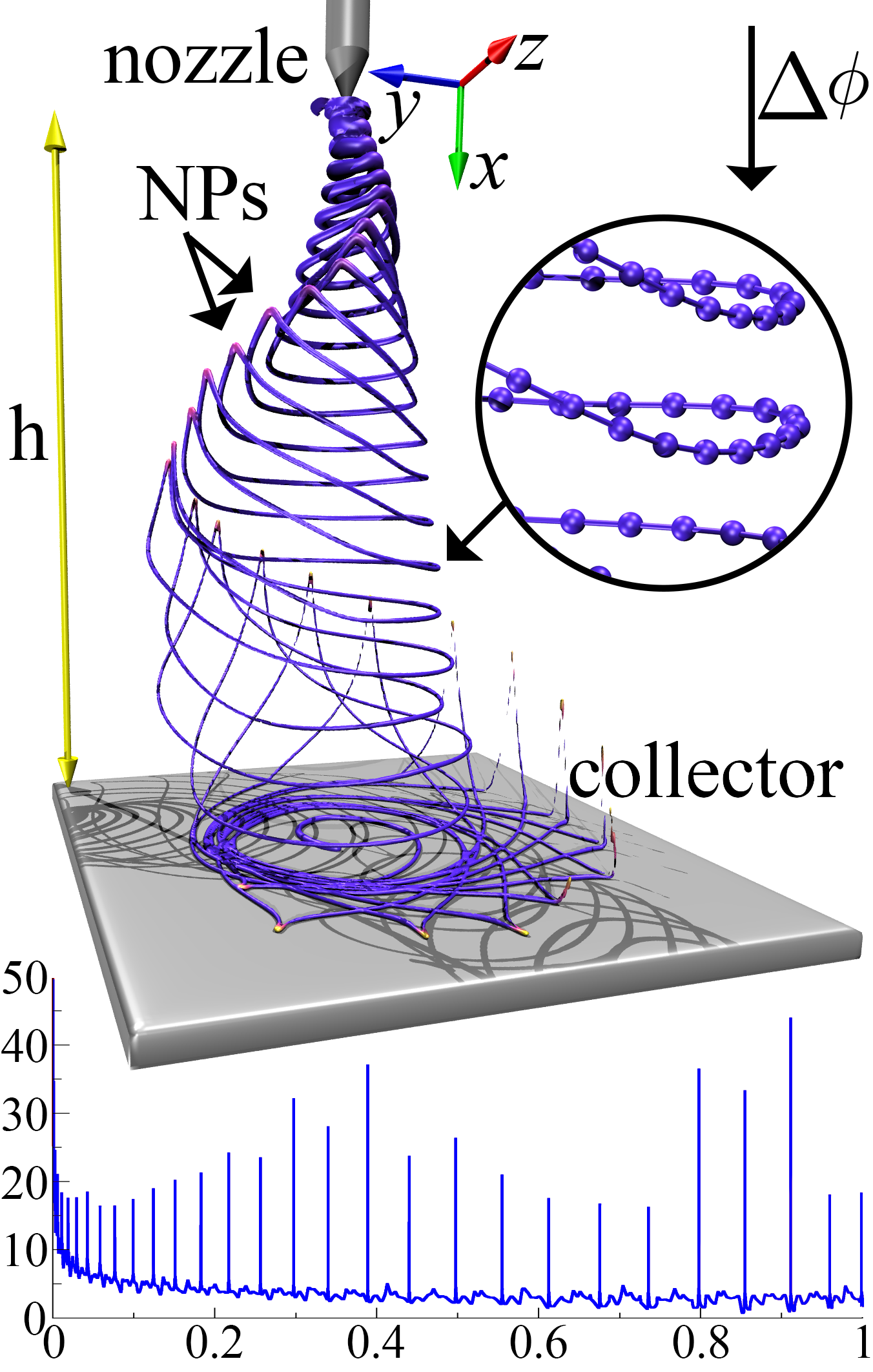}
\par\end{centering}
\caption{On the top, a simulation snapshot illustrating the electrospinning process: a jet is ejected at the nozzle and it is stretched towards a conductive collector
by an external electric potential $\Delta \phi$ imposed between the nozzle
and the collector. The color map used to drawn the jet is reported in Fig. \ref{fig:snaps}.
On the bottom, the decreasing profile of the jet radius $a$ (in $\mu$m) versus 
the curvilinear coordinate $\lambda\in[0,1]$, where $\lambda=0$ denotes 
the nozzle while $\lambda=1$ the last jet element closest to the collector without touching it.}
\label{fig:schema}
\end{figure}

In this Letter, we investigate the effects of nanoparticles (NPs) inserted
in a polymeric liquid bulk on the resulting charged jet dynamics (see Fig. \ref{fig:schema}). 
Interest in this process has been spurred by the possibility of 
tailoring the material composition and the physical properties 
of nanocomposite nanofibers \cite{zhang2014nanoparticles}.
Examples include reinforced yarns with carbon nanotubes \cite{ko2003electrospinning}, 
fluorescent quantum dots embedded in fibers to show suppressed energy 
transfer \cite{li2007electrospinning} or single-photon coupling to optical 
modes transmitted in sub-wavelength waveguides\cite{gaio2016modal}, 
nanodiamonds loaded at high concentrations to obtain coatings for UV 
protection and scratch resistance \cite{behler2009nanodiamond}, application 
of metal NPs to surface-enhanced Raman scattering\cite{zhang2012controlled} 
and to nonvolatile flash memories\cite{chang2013flexible}.

In all aforementioned applications, the polymer component serves as a three-dimensional 
topological network of filaments in which NPs compose distributed functional domains.
However, encoding the spatially-resolved information embedded in the nanofibers requires
a detailed understanding of the way that electrospinning instabilities are modulated 
by the presence of NPs or clusters thereof, which can profoundly affect the ultimate 
jet morphology. 
Despite the major interest in above phenomena, to the best of our
knowledge, numerical investigations of the dynamic behavior
of electrified jets loaded with NPs are still lacking. 
In this Letter we take a first step along this line.

In the present study, we extend the discrete element
model originally introduced by Reneker and co-workers \cite{reneker2000bending},
as discussed in Refs. \cite{lauricella2016dynamic,lauricella2017effects},
and recently implemented in the open source code JETSPIN\cite{lauricella2015jetspin,jetspin}.

Briefly, the jet is discretized into $n$ particle-like elements, representing a 
cylindrical jet segment, each labeled by the discrete index $i=1,..,n$, 
with mass $m_{i}$, charge $q_{i}$ and volume $V_{i}$.
Each jet element is inserted at the nozzle at a mutual distance $l_{i}=l_{step}$ 
(initial length step of discretization), 
from the last bead, and imposing an initial jet volume $V_i=V_{0}$, 
corresponding to an initial jet radius $a_{0}=\sqrt{V_{0}/\pi l_{step}}$. 

Under the effect of the electric potential $\Delta \phi$
 (see Fig. \ref{fig:schema}), the jet elements 
move away from the nozzle undergoing a stretching process
and a consequent decrease of the filament radius $a_{i}$
as a result of the volume conservation.
For the sake of self-consistency, we review the basic model in Supplemental 
Material, while in the following we focus on the original features of the NP modeling.
To this purpose, let us consider the $i$-th jet element with a spherical NP of radius $R$
embedded in the filament. 
The NP is assumed to be frozen in the polymeric bulk, hence following the same dynamic trajectory. 

The mass (tagged by a star) of the cylindrical segment 
of volume $V_{i}$ including the NP is given by:
\begin{equation}
m_{i}^{*}=\left(V_{i}-\frac{4}{3}\pi R^{3}\right)\rho_{l}+\frac{4}{3}\pi R^{3}\rho_{NP},\label{eq:mass-star}
\end{equation}
where $\rho_{l}$ and $\rho_{NP}$ are the mass densities of the viscoelastic
liquid and of the NP, respectively. Note that we are assuming $R<l_{i}$,
so that each NP is confined to a single jet element. 
Further, we assume that only the polymeric bulk of the jet element is deformable. 
As a consequence, the viscoelastic stress imparts a force proportional to the
cross sectional area $\pi {a_{i}^{*}}^{2}$ of the deformable part of the element, whose
effective radius is defined as $a_{i}^{*}=\sqrt{\left(V_{i}-4/3\pi R^{3}\right)/\pi l_{i}}$.

\begin{table}
\begin{centering}
\begin{tabular}{ccccc}
\hline 
$\rho_{l}$  & $\rho_{NP}$  & $\rho_{q}$  & $a_0$ & $h$ \tabularnewline
($\text{g}/\text{cm}^{3}$)  & ($\text{g}/\text{cm}^{3}$)  & ($\text{statC}/\text{cm}^3$)  & ($\text{cm}$) & ($\text{cm}$) \tabularnewline
\hline 
\hline 
0.84  & 10.49  & $8.39\cdot10^{-1}$  & $5\cdot10^{-3}$ & 16 \tabularnewline
\hline 
\end{tabular}
\par\end{centering}
\medskip{}

\begin{centering}
\begin{tabular}{cccc}
\hline 
$\alpha$  & $\eta$  & $Y$  & $\Delta \phi$ \tabularnewline
(g/$\text{s}^{2}$)  & (g/cm s)  & (g/cm $\text{s}^{2}$)  & (statV)  \tabularnewline
\hline 
\hline 
21.1  & 0.2  & $5\cdot10^{4}$  & 30.0  \tabularnewline
\hline 
\end{tabular}
\par\end{centering}
\medskip{}

\caption{Main simulation parameters in cgs units: 
$\rho_{l}$ density of liquid bulk, $\rho_{NP}$ density
of NP, $\rho_{q}$ charge density, $a_0$ initial jet radius at the nozzle,
$h$ distance between the nozzle and the collector,
$\alpha$ surface tension, $\eta$ dynamic viscosity, $Y$ Young's modulus,
$\Delta \phi$ applied voltage. With the exception of 
$\rho_{NP}$ (mass density of Ag), all the parameters were taken from Ref. \cite{lauricella2015jetspin}.}

\label{tab:parameters}
\end{table}

\begin{figure}[ht]
\begin{centering}
\includegraphics[width=0.46\textwidth]{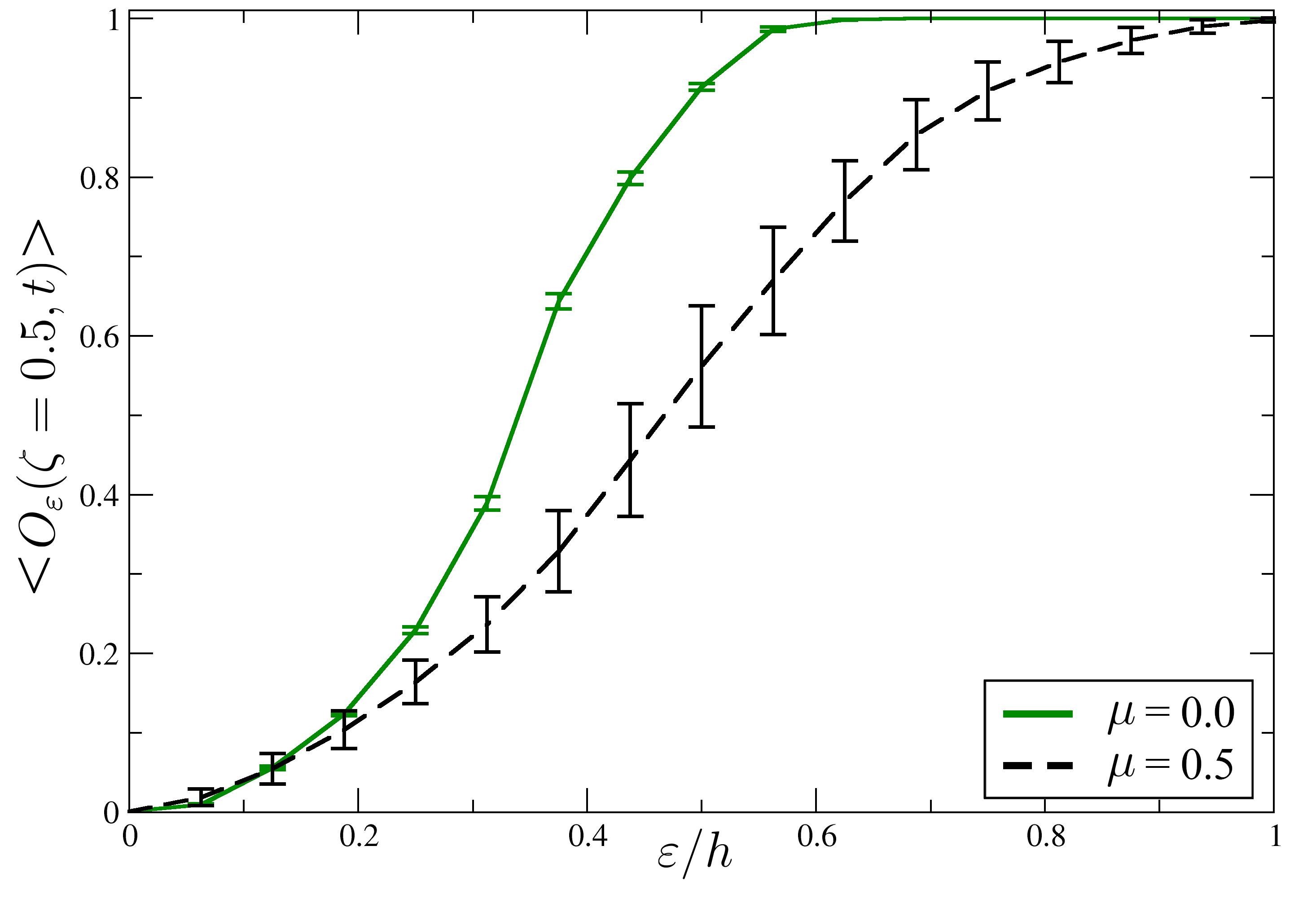}
\par\end{centering}
\caption{Mean value of SOP computed in the stationary regime 
($\zeta=0.5$) as function of ratio $\varepsilon/h$.
With $\varepsilon=8$, namely half distance
between the nozzle and the collector, a neat separation between 
the cases $\mu=0$ and $\mu=0.5$, is well visible.
}
\label{fig:meansop}
\end{figure}

As a reference case, we take the simulation in absence of NPs (unloaded jet), with 
the simulation parameters assessed in Ref. \cite{lauricella2015jetspin}  (see Table \ref{tab:parameters}). 
The discretization step length is set to $0.02$ cm, 
in line with typical values reported in literature  
\cite{reneker2000bending,kowalewski2005experiments,thompson2007effects,
sun2010three,carroll2011discretized,lauricella2015jetspin},
while the initial jet radius is set to $a_0=0.005$ cm.

We investigate seven different values of the particle radius, all other simulation parameters being kept fixed. 
We consider particles with mass density $\rho_{NP}=10.49 \text{g}/\text{cm}^{3}$ (Ag),
while the mass bulk density is set to  $\rho_{l}$ is $0.84  \text{g}/\text{cm}^{3}$. 
According to Eq \ref{eq:mass-star}, we choose the 
NP radius $R$ so as to prescribe specific values of the {\it excess mass ratio}
$\mu=(m^{*}/m)-1$
at the injection point.
In the above, $m^{*}$ and $m$ denote the
inserted masses with and without NPs, respectively. 

In all simulations, we load one jet element every five.
Since each element represents a cylindric jet segment of length $0.02$ cm at the nozzle, we simulate
a liquid jet with a series of NPs regularly spaced at a distance $0.1$ cm. 

\begin{figure}[ht]
\begin{centering}
\includegraphics[width=0.46\textwidth]{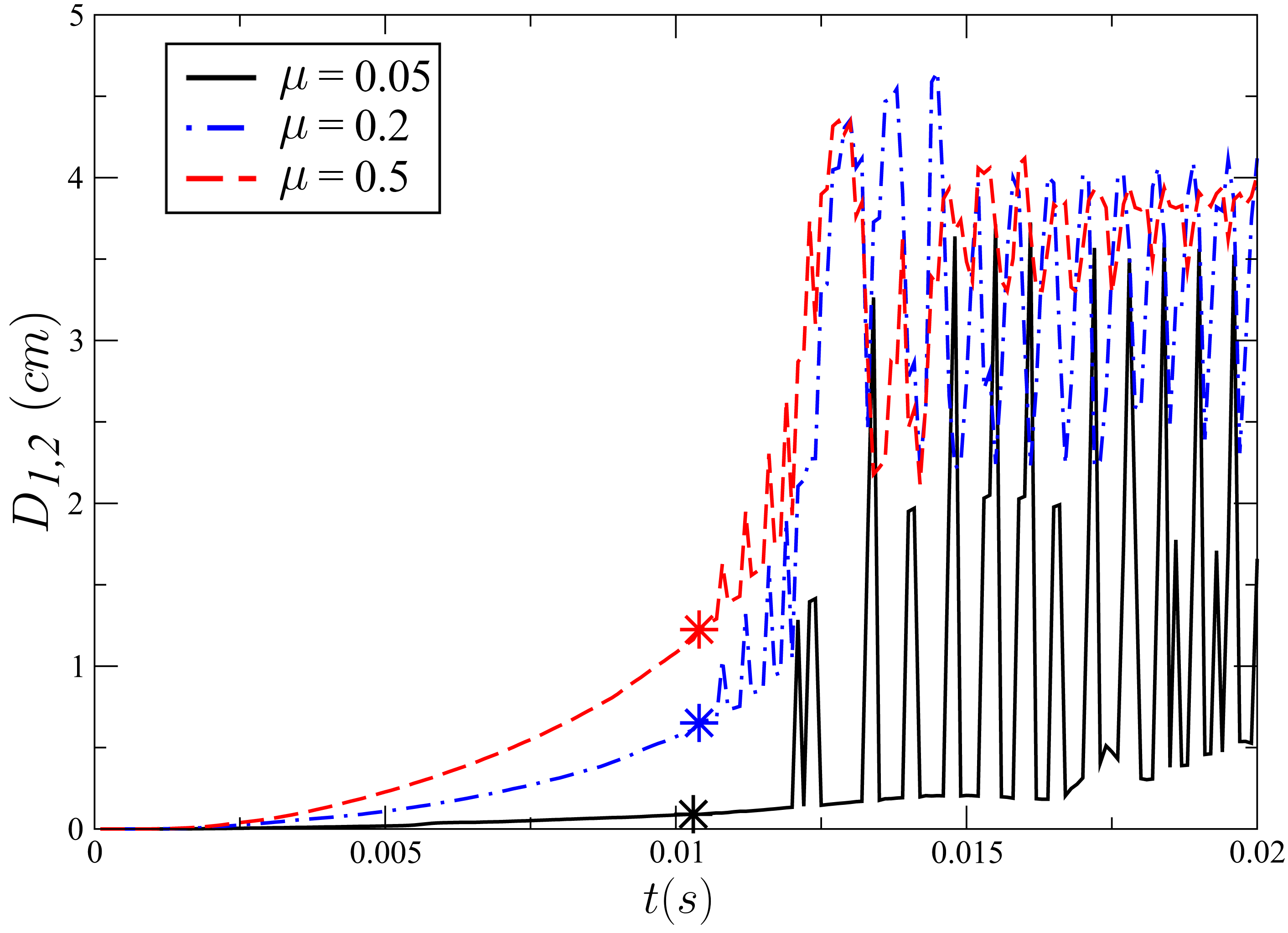}
\par\end{centering}
\caption{Dynamic evolution of $D_{0,\mu}$ versus time for three
cases $\mu=0.05$, $\mu=0.2$ and $\mu=0.5$.
The initial stage is characterized by a near-quadratic growth of the distance,
with a mass-dependent acceleration. Upon hitting the collector, the distance 
undergoes oscillations around a non-zero mean. Both the amplitude and the non-zero
mean are increasing functions of the excess-mass.
}
\label{fig:s_ax}
\end{figure}

As a synthetic indicator of the departure between the jet 
path with and without NP, we introduce the global "distance" 
between two jet configurations: 
\begin{equation}
D_{1,2}(t)= \int_{0}^1 |\vec{r}_2(\lambda,t)-\vec{r}_1(\lambda,t)| \; d \lambda,
\end{equation}
where $\lambda\in\left[0,1\right]$ is  the curvilinear coordinate with $\lambda=0\,(1)$ denoting the nozzle (collector), respectively. 
This distance measures the cumulative point-by-point deviation  
between homologue (same $\lambda$) elements of the paths $1$ and $2$.
The condition $D_{1,2}/h \ll 1$ denotes close-by configurations, as
they occur in the unloaded limit $\mu \to 0$.

It is also convenient to define a local 
{\it self-overlap parameter} (SOP) as follows: 

\begin{equation}
O_{\varepsilon} \left( \zeta, t \right)= \int_{0}^1 
H(\varepsilon - d(\lambda , \zeta))  \; d \lambda
\label{eq:obs-fingerprint}
\end{equation}
where $H$ is the Heaviside step function and $\zeta\in\left[0,1\right]$
specifies a generic point along the jet path, and 
$d(\lambda , \zeta) \equiv |\vec{r}(\lambda)-\vec{r}(\zeta)|$ is the local distance
along the path.
In the present context, low (high) SOP indicate high (low) stretch of the filament,
which is qualitatively associated with local unstable (stable) behaviour.
By definition, $O \to 0$ in the limit $\varepsilon \to 0$ and
$O \to 1$ in the limit $\varepsilon \to \infty$, whatever the path morphology. 
As shown in Fig. \ref{fig:meansop}, SOP works best away from both limits,  
namely for $\epsilon$ smaller than a typical macroscale, say the 
path length $L \sim h$, and larger than the smallest 
lengthscale, say, the helical pitch.

In Fig. \ref{fig:s_ax} we report the observable $D_{0,\mu}$, 
versus time, computed for the unloaded case $\mu=0$ 
versus loaded ones, $\mu>0$.
From this figure, we observe that the distance between the different 
configurations scales quadratically in time, indicating a free-fall-like 
regime before hitting the collector.
As expected, the acceleration is a decreasing function of the 
mass excess parameter $\mu$.
Upon hitting the collector, the distance shows periodic oscillations
around a non-zero mean, of the order of a few centimeters, i.e. smaller
but non-negligible as compared to nozzle-collector distance $h=16$ cm.
As we shall detail shortly, these oscillations are interpreted as
the signature of local inverted V-shape upturns of the 
trajectory, caused by the excess mass of the NPs, which tends to slow
down the loaded segments relative to the unloaded ones.

\begin{figure}[ht]
\begin{centering}
\includegraphics[width=0.46\textwidth]{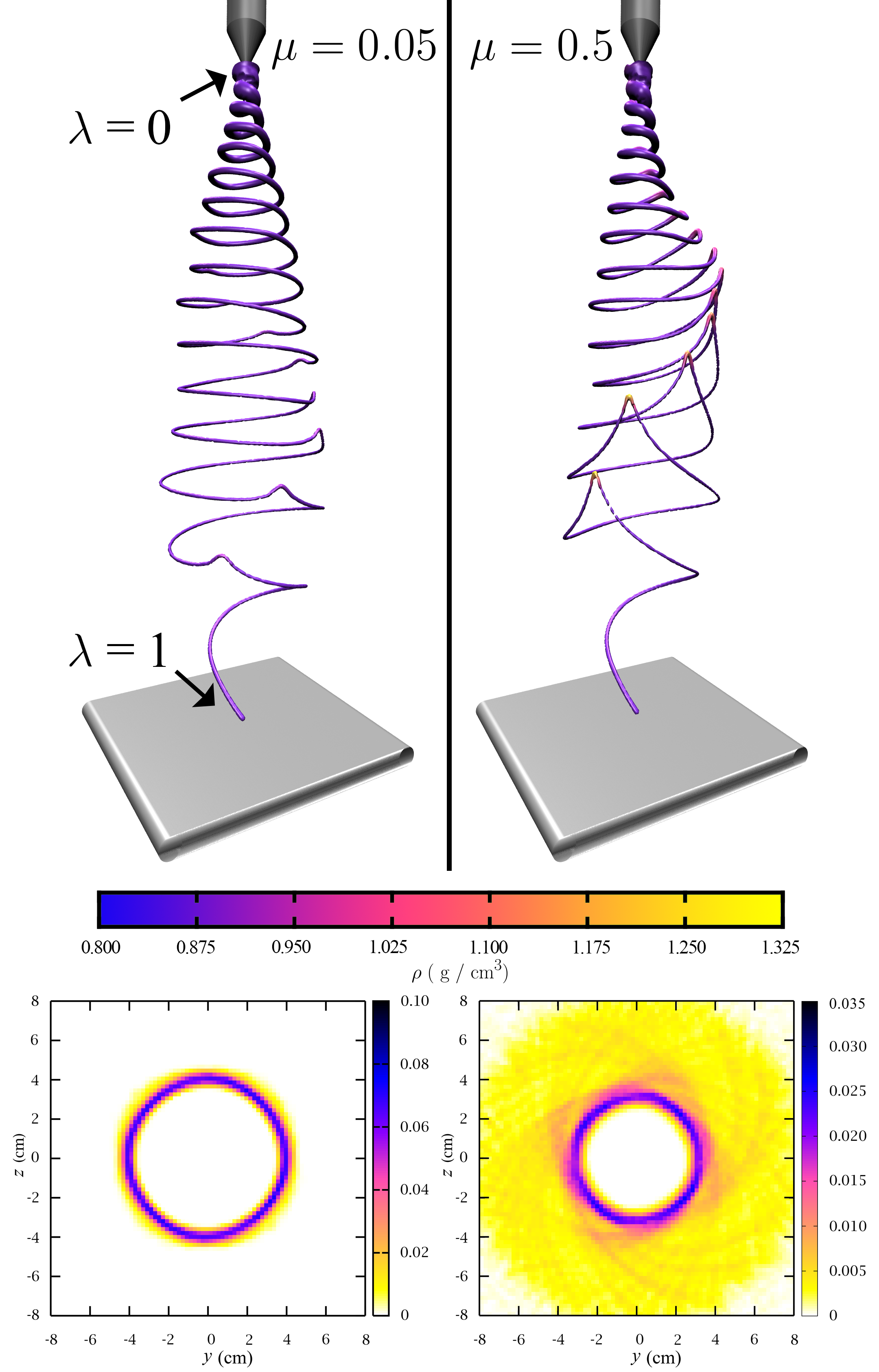}
\par\end{centering}
\caption{On the top panel, two snapshots of the cases $\mu=0.05$ and 
$\mu=0.5$, taken at $t=0.01$ seconds alongside the color map, showing the colors used to draw
the jet according to the corresponding mass density value $\rho$. 
On the bottom panel, the normalized 2D maps computed over the coordinates $y$ and $z$ of the collector
for the two aforementioned cases collected over all the simulation runs with  
their respective color palettes defining the probability that
a jet bead hits the collector at coordinates $y$ and $z$.}
\label{fig:snaps}
\end{figure}

To gain further insight into the local morphology of the
loaded jet, we inspect two snapshots taken at time
$0.01$ seconds for the cases $\mu=0.05$ and $\mu=0.5$.
As anticipated, this figure clearly shows that the
slowing-down is due to the NPs, resulting from local inverted-V shaped upturns
of the path configuration
(higher mass density highlighted in yellow in Fig. \ref{fig:snaps}).

\begin{figure}[ht]
\begin{centering}
\includegraphics[width=0.46\textwidth]{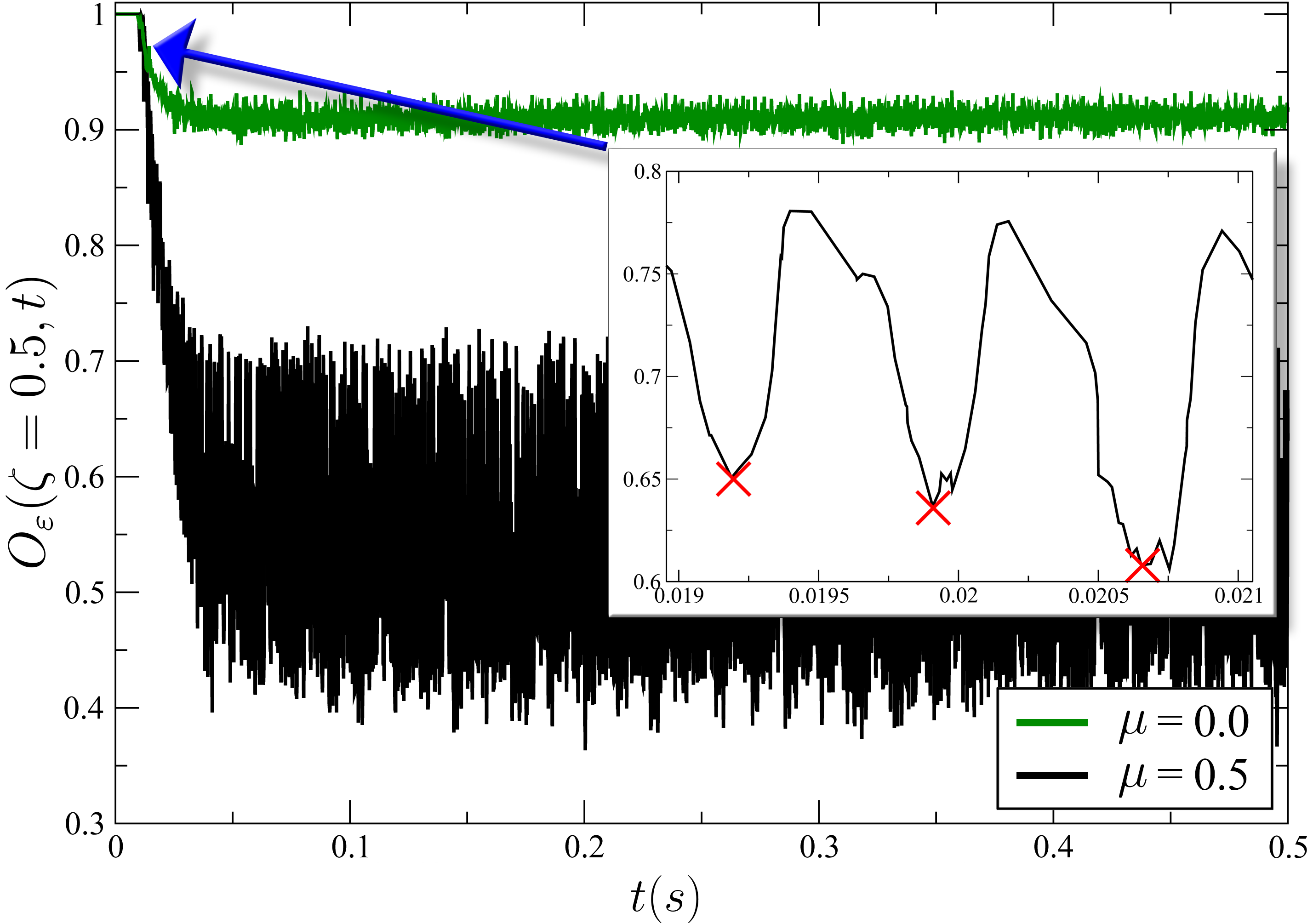}
\par\end{centering}
\caption{Dynamic evolution of $O_{\varepsilon}\left(\zeta=0.5,t\right)$ versus time
for the two simulations with $\mu=0.0$ and $\mu=0.5$.
In the inset we enlarge the dynamic for the case $\mu=0.5$ showing in red cross symbols 
whenever a NP is embedded in the jet point  $\zeta=0.5$.}
\label{fig:fingerprint-t}
\end{figure}

In Fig. \ref{fig:fingerprint-t}, we report the time evolution
of SOP for the cases $\mu=0$ and $\mu=0.5$.
From this figure, we observe that both cases show a short decay to
a steady state, with periodic oscillations on top of it. 
The case $\mu=0.5$ exhibits a much smaller SOP than $\mu=0$, and much larger
oscillations all along, which is again a dynamic signature of the local upturns.
Note that Fig. \ref{fig:fingerprint-t} refers to the 
mid-point $\zeta=0.5$, with a distance threshold $\varepsilon=8$ cm, i.e
half the nozzle-collector separation, which is the best value
to detect departures along the jet filament, as reported in Fig \ref{fig:meansop}.
It would be interesting to explore whether such collective observables 
could be monitored experimentally. This could be realized by means of recently 
proposed fast jet imaging using semi-transparent diffusing screens \cite{montinaro2015sub}, 
properly combined with stereoscopic reconstruction techniques \cite{michailidis2014high}.
 
\begin{figure}[ht]
\begin{centering}
\includegraphics[width=0.46\textwidth]{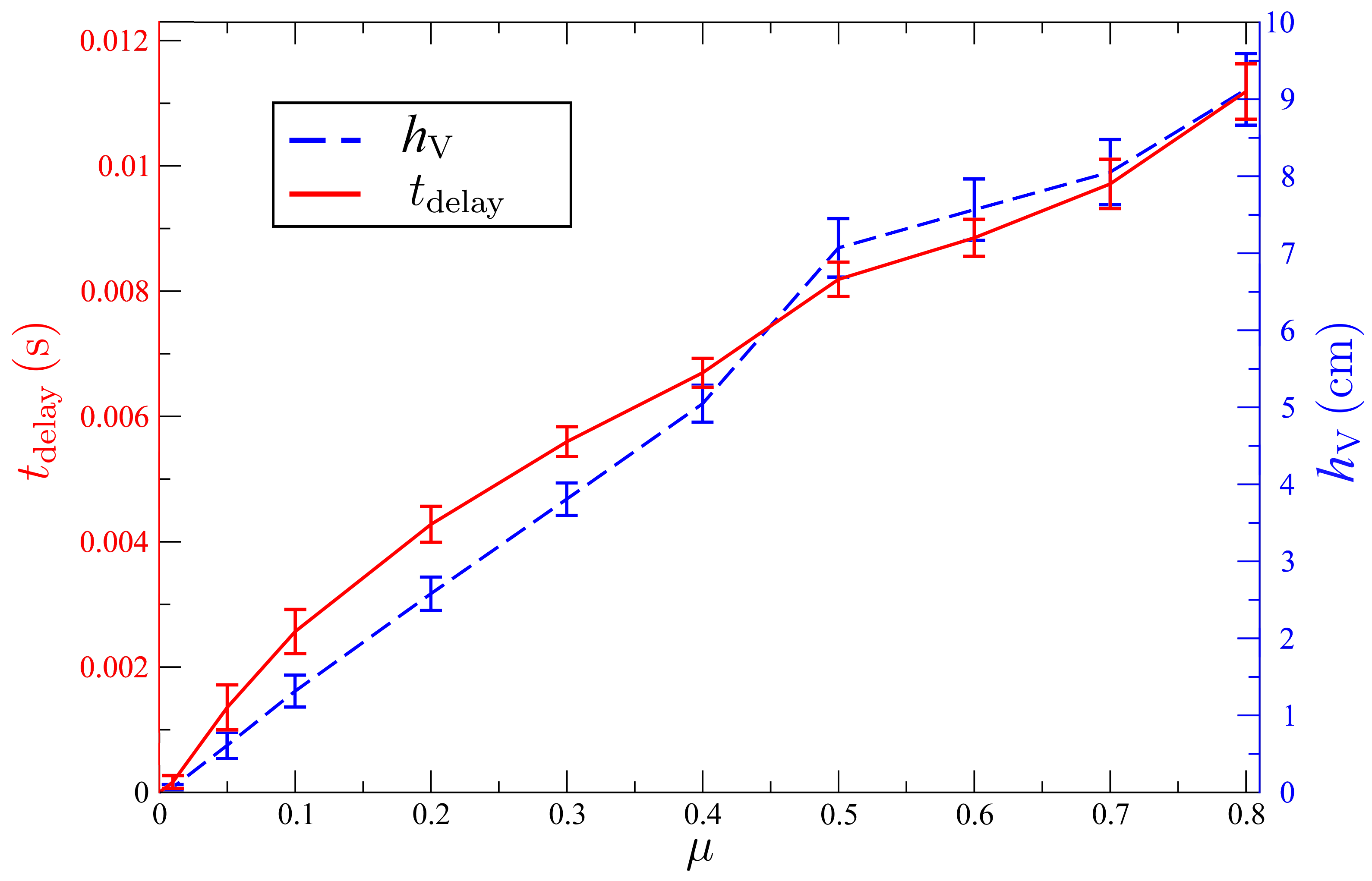}
\par\end{centering}
\caption{Averaged values of the V shape height $h_{\text{V}}$ (dashed blue line)
and the delay time $t_{\text{delay}}$ (continuous red line) measured at the
collector, as soon as the inverted V shape touches the conductive plate on both
its sides. On the abscissas, we report the ratio $\mu$ of the masses related to
the jet element with and without NP.}
\label{fig:vshape}
\end{figure}

\begin{figure}[!ht]
\begin{centering}
\textit{\includegraphics[width=0.46\textwidth]{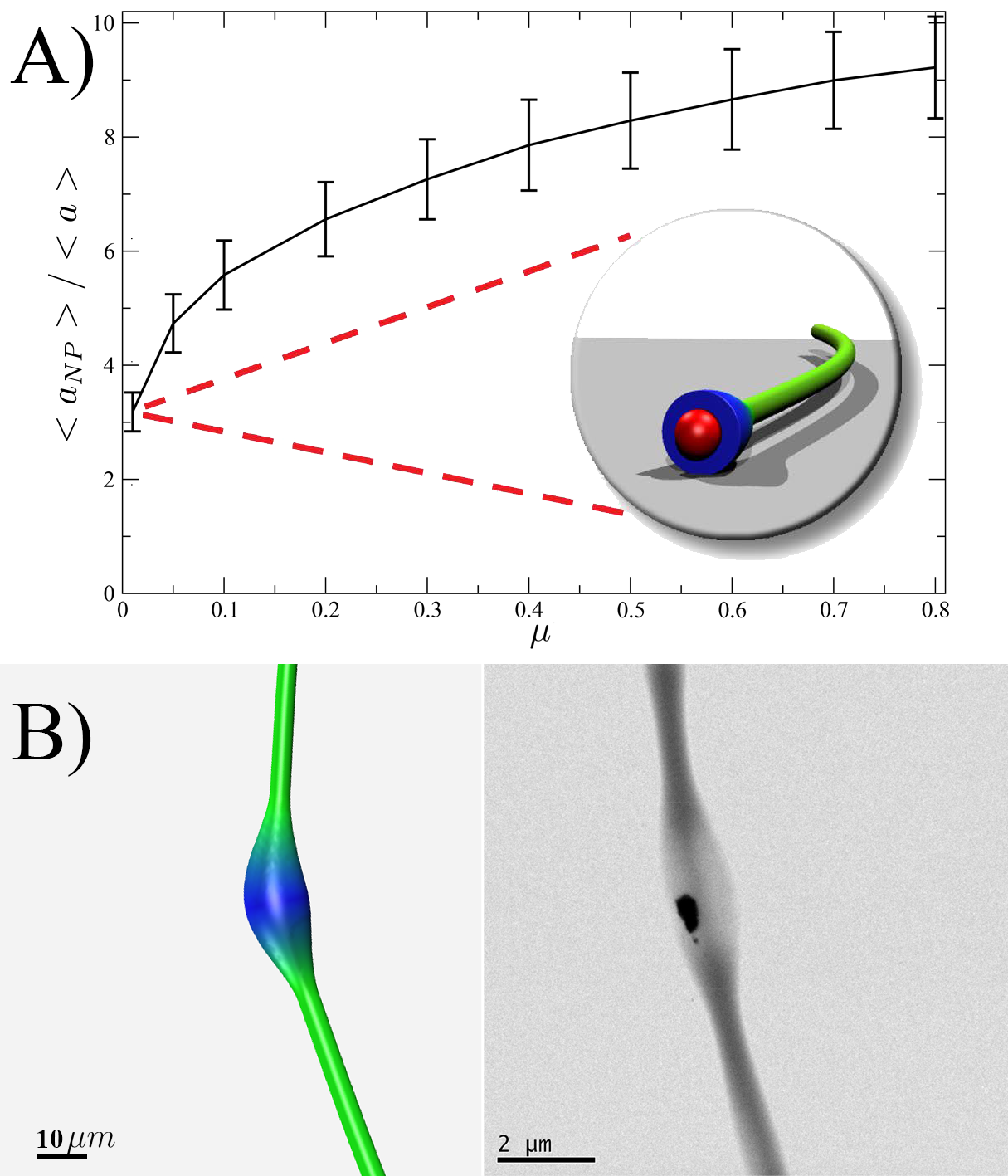}}
\par\end{centering}
\caption{Panel A: Ratio $<a_{NP}>/<a>$ between average jet radii of fiber elements with
and without NP for all the cases under investigation. 
On the abscissas, we report the ratio $\mu$ of the masses related to
the jet element with and without NP as inserted at the nozzle.
In the inset, the cross section of the fiber containing a NP (drawn in red) is shown for the simulation performed at $\mu=0.01$.
Panel B: Left: a snapshot of a fiber segment containing a NP with $\mu=0.01$ (the same
of the above inset although from a different perspective).
Right: an experimental snapshot of an electrospun, particle-embedding polymer 
filament.}
\label{fig:cross-sec}
\end{figure}

The changes in the jet morphology due to embedded NPs 
bears a significant impact also on the final deposition pattern of 
the nanofibers at the collector.

In Fig. \ref{fig:snaps}, we show the in-plane probability of a 
jet bead hitting the collector around the coordinates
$y$ and $z$ (the plate being perpendicular to $x$ direction)
for the cases $\mu=0$ and $\mu=0.5$. 
From Fig. \ref{fig:snaps}, we observe that the presence of the heaviest 
NPs shifts part of the deposited filament towards an
outer shell, more than 8 cm away from the origin.
The shift is also evident in Fig. \ref{fig:schema}, which reports
a snapshot of the case $\mu=0.5$, with NPs (drawn in yellow) of the deposited
fiber occupying the outermost area of the collector. 

The effect is mainly due to the longer flight time of the jet 
segments containing NPs, which move at lower speed.
The delay flight-time allows the filament to take and
outward radial shift, while unloaded jet segments
hit the collector along a well defined circle, with no radial 
shift due to the disturbance of NPs.

In Fig. \ref{fig:vshape}, we report the delay flight-time as a function of $\mu$. 
A monotonically increasing trend is clearly visible, as 
a result of the V-shaped upturns due to the localized NPs.
The same figure also reports the height of the V-shaped upturn, $h_V$, defined
as the height of the triangle whose upper vertex is defined by the NP location, while
the two lower vertices coincide with the touch-down locations on the collector.
As expected, a similar trend as $t_{delay}$ is observed. 
In either cases, no abrupt change is observed, indicating no evidence of "critical"
thresholds of the $\mu$ parameter.

In Fig. \ref{fig:cross-sec}, we report the average fiber
radius $<a>$, measured for both loaded ($a_{\text{NP}}$) and unloaded ($a$),  jet elements
for the cases under investigation, where brackets denote time average over the simulation time.

The figure shows a significant increase of the fiber
radius as a function of $\mu$, which is sizeable already at $\mu=0.01$.
In particular, we observe a rapid variation in the jet radius profile 
versus the curvilinear coordinate, resulting in a varicose-like shape of the filament.
This delivers a fiber radius at the collector three times larger than the unloaded case,
providing a concrete example of varicosity phenomena affecting the size
of electrospun fibres.
Varicosity effects are indeed detected in ongoing electrospinning 
experiments (panel B of Fig. \ref{fig:cross-sec}).
This figure shows a typical varicose profile, due to the insertion
of a NP cluster with radius $R=0.4$ microns, leading to a local
jet radius $a \sim 0.8$ microns, i.e. a factor two enhancement as compared
to the unloaded case.
The qualitative agreement between simulation (left) and experiment (right)
is intriguing, and makes a subject of decided interest for future research.

\section*{Acknowledgments}
This research has been funded from the European Research Council under 
the European Union's Seventh Framework Programme (FP/2007-2013)/ERC Grant 
Agreement n. 306357 (ERC Starting Grant ''NANO-JETS'').
R. Manco, C. Nobile and P.D. Cozzoli are gratefully acknowledged 
for the experimental part of Fig. \ref{fig:cross-sec}.

\newpage{}

\end{document}